\documentclass[fleqn,usenatbib]{mnras}

\usepackage{newtxtext,newtxmath}
\usepackage{footnote}

\usepackage[T1]{fontenc}
\usepackage{ae,aecompl}

\usepackage{graphicx}	
\usepackage{amsmath}	
\usepackage{amssymb}	

\title[WASP-177, WASP-181 and WASP-183]{Three Hot-Jupiters on the upper edge of the mass-radius distribution: WASP-177, WASP-181 and WASP-183}

\author[O.D.Turner et al.]{
Oliver D. Turner,$^{1}$\thanks{E-mail: oliver.turner@unige.ch}
D.R. Anderson,$^{2}$
K. Barkaoui,$^{3,4}$,
F. Bouchy,$^{1}$
Z. Benkhaldoun$^{4}$
\newauthor
D.J.A. Brown,$^{5,6}$
A. Burdanov,$^{3}$
A. Collier Cameron,$^{7}$
E. Ducrot,$^{3}$
M. Gillon,$^{3}$
\newauthor
C. Hellier,$^{2}$
E. Jehin,$^{3}$
M. Lendl,$^{8,1}$
P.F.L. Maxted,$^{2}$
L.D. Nielsen,$^{1}$
F. Pepe,$^{1}$
\newauthor
D. Pollacco,$^{5,6}$
F.J. Pozuelos,$^{3}$
D. Queloz,$^{1,9}$
D. S\'egransan,$^{1}$
B. Smalley,$^{2}$
\newauthor
A.H.M.J. Triaud,$^{10}$
S. Udry$^{1}$,
and R.G. West$^{5,6}$
\\
$^{1}$Observatoire de Gen\`eve, Universit\'e de Gen\`eve, 51 Chemin des Maillettes, 1290 Sauverny, Switzerland\\
$^{2}$Astrophysics Group, Keele University, Staffordshire ST5 5BG, UK\\
$^{3}$Space sciences, Technologies and Astrophysics Research (STAR) Institute, Universit\'e de Li\`ege, Li\`ege 1, Belgium\\
$^{4}$Oukaimeden Observatory, High Energy Physics and Astrophysics Laboratory, Cadi Ayyad University, Marrakech, Morocco\\
$^{5}$Department of Physics, University of Warwick, Coventry CV4 7AL, UK\\
$^{6}$Centre for Exoplanets and Habitability, University of Warwick, Gibbet Hill Road, Coventry CV4 7AL, UK\\
$^{7}$SUPA, School of Physics and Astronomy, University of St. Andrews, North Haugh, Fife KY16 9SS, UK\\
$^{8}$Space Research Institute, Austrian Academy of Sciences, Schmiedlstr. 6, A-8042 Graz, Austria\\
$^{9}$Cavendish Laboratory, J J Thomson Avenue, Cambridge CB3 0HE, UK\\
$^{10}$School of Physics \& Astronomy, University of Birmingham, Edgbaston, Birmingham, B15 2TT, UK\\
}

\date{Accepted XXX. Received YYY; in original form ZZZ}

\pubyear{2018}

\begin{document}
\label{firstpage}
\pagerange{\pageref{firstpage}--\pageref{lastpage}}
\maketitle

\begin{abstract}
We present the discovery of 3 transiting planets from the WASP survey, two hot-Jupiters: WASP-177\,b ($\sim$0.5 M$_{\rm Jup}$, $\sim$1.6 R$_{\rm Jup}$) in a 3.07-d orbit of a $V = 12.6$ K2 star, WASP-183\,b ($\sim$0.5 M$_{\rm Jup}$, $\sim$1.5 R$_{\rm Jup}$) in a 4.11-d orbit of a $V = 12.8$ G9/K0 star; and one hot-Saturn planet WASP-181\,b ($\sim$0.3 M$_{\rm Jup}$, $\sim$1.2 R$_{\rm Jup}$) in a 4.52-d orbit of a $V = 12.9$ G2 star. Each planet is close to the upper bound of mass-radius space and has a scaled semi-major axis, $a/R_{*}$, between 9.6 and 12.1. These lie in the transition between systems that tend to be in orbits that are well aligned with their host-star's spin and those that show a higher dispersion.
\end{abstract}




\section{Introduction}

Since the beginning of the project the Wide Angle Search for Planets (WASP; \citealt{2006PASP..118.1407P}) survey has discovered nearly 190 transiting, close-in, giant exoplanets. As they transit their host stars their bulk properties, mass and radius, can be determined relatively easily. Their transits allow for deeper characterisation that has led to the discovery of multiple chemical and molecular species in their atmospheres \citep{2013MNRAS.436L..35B,2013A&A...554A..82D,2017A&A...602A..36W,2018Natur.560..453H} and the observation of planetary winds \citep{2016ApJ...817..106B}. 

Close-in exoplanets can also provide information on the formation and migration mechanisms of solar systems. It is expected that hot-Jupiter exoplanets initially form much further from their stars than where we detect them today. Therefore some mechanism must cause this migration. There are two proposed pathways, high eccentricity migration or disk migration. In the former some mechanism e.g. Kozai cycles \citep{2003ApJ...589..605W,2013apf..book.....A} or planet-planet scattering \citep{1996Sci...274..954R,1996Natur.384..619W},  forces the cold Jupiter into a highly eccentric orbit which then is tidally circularised via interaction with the star. During this kind of migration it is possible for the planet orbital axis to become mis-aligned with the stellar spin axis \citep{2007ApJ...669.1298F}. 
In the latter mechanism the planet loses angular momentum via interaction with the stellar disk during formation and migrates inward \citep{1980ApJ...241..425G}.
 This is expected to preserve the initial spin-orbit alignment \citep{2009ApJ...705.1575M}, though work is being done to investigate the production of mis-aligned planets due to inclined protoplanetary discs \citep{2017MNRAS.471.2334X}. 
 
The alignment between the stellar rotation axis and planet orbit can be investigated with the Rossiter-M$^{\rm c}$Laughlin (RM) technique (\citealt{1924ApJ....60...15R,1924ApJ....60...22M,2010A&A...524A..25T}, etc.). These observations have shown a general trend for systems orbiting cool stars (with $T_{\rm eff} < 6250$K; \citealt{2012ApJ...757...18A,2015ApJ...800L...9A}) to be more well aligned than systems orbiting hotter stars. Tides are also expected to play a role. In cool star systems, those with smaller scaled semi-major axes, $a/R_{*}$, tend to be more often well aligned than those with larger $a/R_{*}$. Though this picture is far from clear as there seems to be evidence for the hot/cool alignment disparity holding even for systems with large separations or low mass planets meaning tidal effects should be minimal (\citealt{2015ApJ...801....3M}) casting tidal realignment into doubt (see also the discussion of  \citealt{2017AJ....153..205D}).

In this paper we present the discovery of three systems at the upper edge of the mass-radius envelop of hot-giants that could be useful probes of tidal re-alignment.

\section{Observations}

Each of these planets was initially flagged as a candidate in data taken with both WASP arrays located at Roque de los Muchachos Observatory on La Palma and at the South African Astronomical Observatory (SAAO). The data were searched for periodic signals using a BLS method as per \cite{2006MNRAS.373..799C,2007MNRAS.380.1230C}. The survey itself is described in more detail by \cite{2006PASP..118.1407P}. 

In order to confirm the planetary nature of the signals radial velocity (RV) data were obtained with the CORALIE spectrograph on the 1.2-m Swiss telescope at La Silla, Chile \citep{2000A&A...354...99Q}. Additional photometry was acquired using EulerCam (\citealt{2012A&A...544A..72L}, also on the 1.2-m Swiss) and the two 0.6-m TRAPPIST telescopes \citep{2011A&A...533A..88G,2011Msngr.145....2J}, based at La Silla and Oukaimeden Observatory in Morocco \citep{Gillon2017Natur,Barkaoui2018}.

Due to the low masses of WASP-181\,b and WASP-183\,b, we also acquired HARPS data\footnote{These observations were made as part of the programs Anderson:0100.C-0847(A) and Nielsen:0102.C-0414(A).}. These observations are summarised in Table \ref{tab:obs}. The TRAPPIST data from 2018-08-13 contain a meridian flip at BJD = 2458344.5639. During analysis the data were partitioned at this point and modeled as two datasets.

\begin{figure}
	\includegraphics[width=\columnwidth]{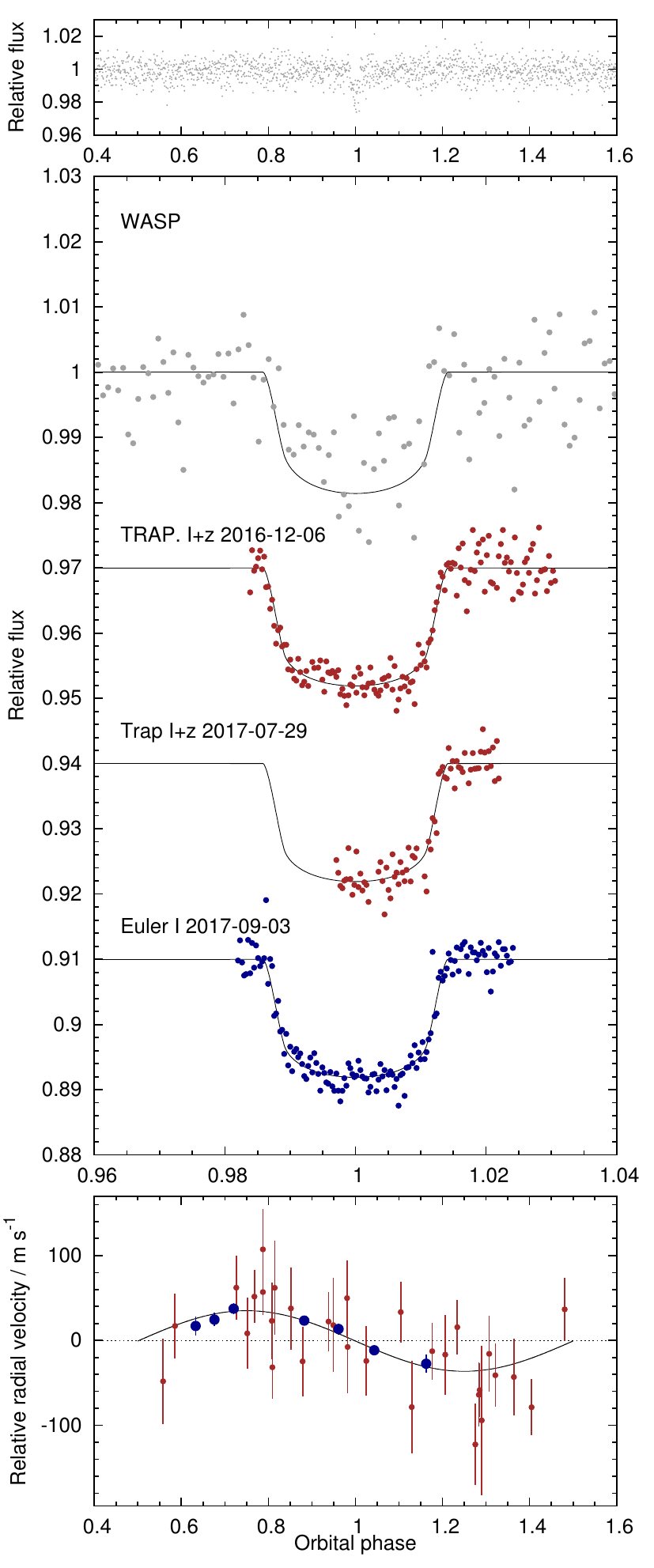}
    \caption{As for Fig.\ref{fig:W177-phot} for the WASP-181 system. CORALIE data in bottom figure are small (red) while HARPS data are larger (blue) symbols.}
    \label{fig:W181-phot}
\end{figure}

\begin{figure}
	\includegraphics[width=\columnwidth]{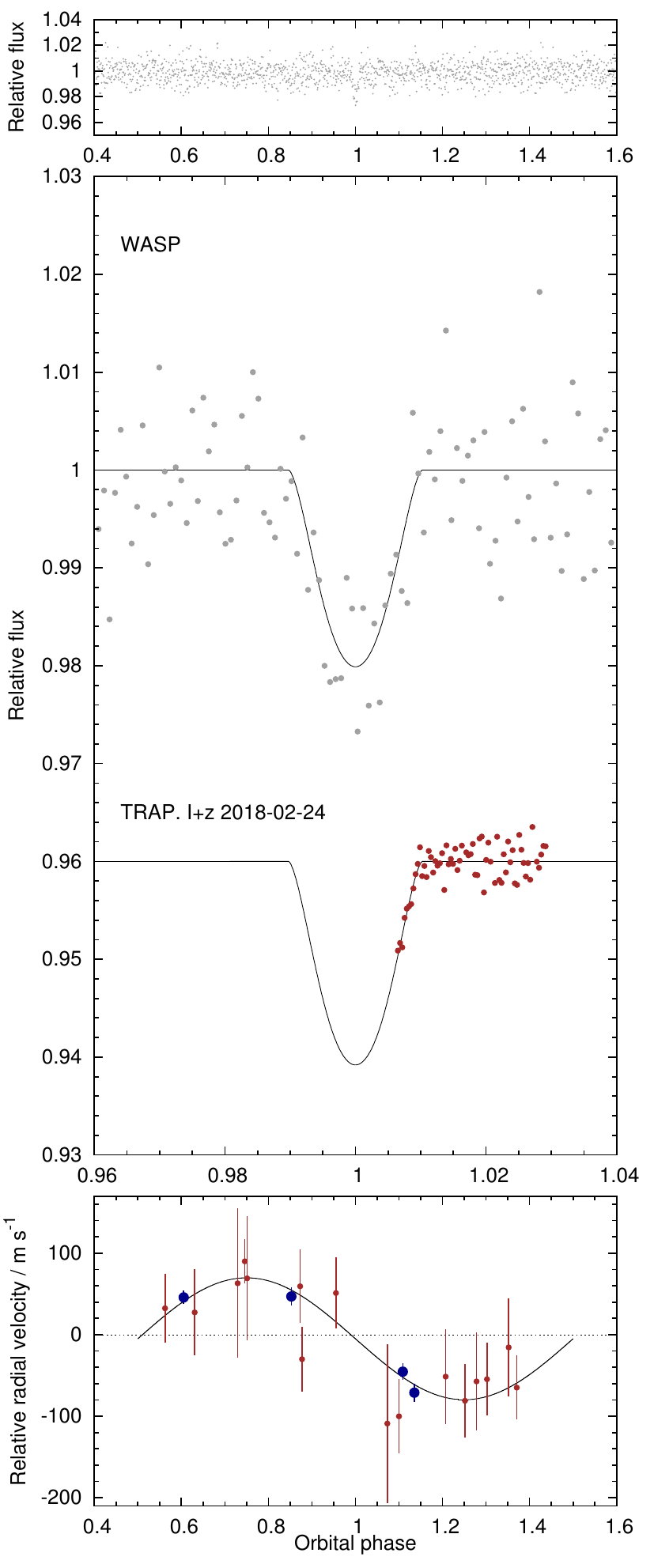}
    \caption{As for Fig.\ref{fig:W181-phot} for the WASP-183 system.}
    \label{fig:W183-phot}
\end{figure}

\begin{figure}
	\includegraphics[width=\columnwidth]{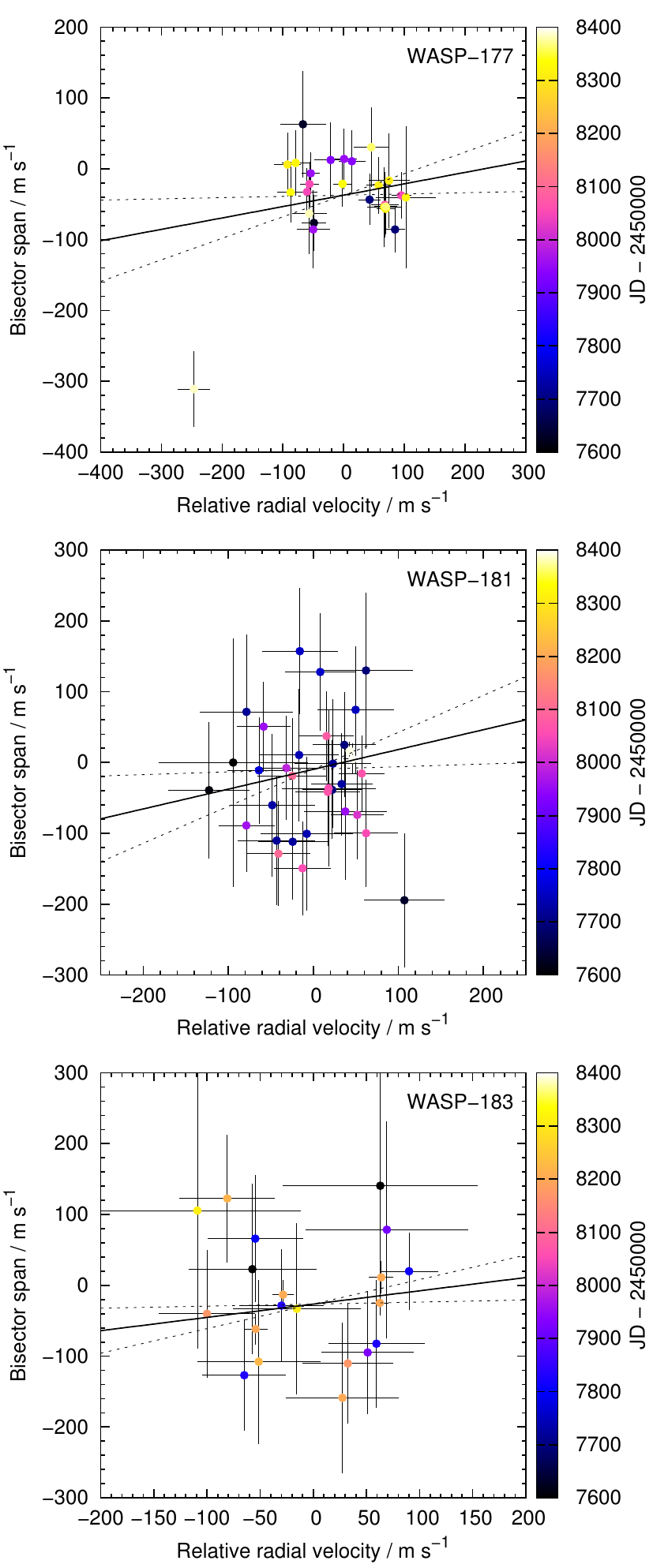}
    \caption{Radial velocity measurements plotted against line bisector spans. There is no strong correlation between the two, thus ruling out transit mimics. Solid lines are the linear best fit to the data. The dotted lines show the 1$\sigma$ uncertainty limits on the fit. }
    \label{fig:bisectors}
\end{figure}

Figures \ref{fig:W177-phot}, \ref{fig:W181-phot} and \ref{fig:W183-phot} show the phase folded discovery and follow-up data. The RVs exhibit signals in phase with those found in the transit data and are consistent with companion objects of planetary mass. We checked for correlation between the RV variation and the bisector spans, see Fig.\ref{fig:bisectors}. We find no strong correlation and so further exclude the possibility that these objects are transit mimics. 

\begin{table}
\begin{minipage}{\textwidth}
\caption{Observations of WASP-177, WASP-181 and WASP-183.}
\label{tab:obs}
\begin{tabular}{llr}\hline
Date  & Source & N.Obs. \\
 & & Filter \\
\hline 
\multicolumn{1}{l}{WASP-177} & & \\
2008 Jul--2010 Oct & WASP (North) & \multicolumn{1}{r}{16\,169}  \\
2008 Jun--2009 Oct & WASP (South) & \multicolumn{1}{r}{10\,825}  \\
2016 Aug--2018 Sep & CORALIE &  \multicolumn{1}{r}{26}  \\
2017 Jul 25 & TRAPPIST-North & I+z \\
2017 Oct 19 & EulerCam & B \\
2018 Jul 13 & EulerCam & V \\
2018 Aug 13 & TRAPPIST-North\footnote{Meridian flip at BJD 2458344.5639.} & I+z \\
\hline 
\multicolumn{1}{l}{WASP-181} & \\
2008 Sep--2010 Dec & WASP (North)& \multicolumn{1}{r}{12\,938}    \\
2008 Jul--2009 Aug & WASP (South)& \multicolumn{1}{r}{9\,059}    \\
2016 Jan--2017 Dec & CORALIE &  \multicolumn{1}{r}{31}    \\
2018 Oct--2019 Jan & HARPS &  \multicolumn{1}{r}{7}    \\
2016 Dec 06 & TRAPPIST-South & I+z \\
2017 Jul 29 & TRAPPIST-North & I+z \\
2017 Sep 03 & EulerCam & I$_{\rm c}$ \\
\\
\hline 
\multicolumn{1}{l}{WASP-183} & \\
2008 Feb--2011 Mar & WASP (North)& \multicolumn{1}{r}{13\,733}   \\
2009 Jan--2010 May & WASP (South)& \multicolumn{1}{r}{10\,789}   \\
2015 May--2018 Jul & CORALIE &  \multicolumn{1}{r}{16}   \\
2018 Mar & HARPS &  \multicolumn{1}{r}{4}    \\
2018 Feb 24 & TRAPPIST-North & I+z \\
\hline 
\end{tabular}
\end{minipage}
\end{table}


\section{Analysis}

\subsection{Stellar Parameters}

To obtain the stellar parameters effective temperature, $T_{\rm eff}$, metallicity, [Fe/H], and surface gravity, $\log g$, we followed the method of \cite{2018MNRAS.475.1809G,2018A&A...615L..13G} using iSpec \citep{2014A&A...569A.111B}. To do this we corrected each spectrum for the computed RV shift, cleaned them of cosmic ray strikes and convolved them to a spectral resolution, $R$, of $47\,000$. Then, ignoring areas typically affected by telluric lines we used the synthetic spectral fitting technique to derive the stellar parameters. Via iSpec we used SPECTRUM \citep{1994AJ....107..742G} as the radiative transfer code with atomic data from VALD \citep{2011BaltA..20..503K}, a line selection based on a $R\sim 47\,000$ solar spectrum \citep{2016csss.confE..22B,2017hsa9.conf..334B} and the MARCS model atmospheres \cite{2008A&A...486..951G} in the wavelength range 480- to 680-nm.  We increased the uncertainties in these parameters by adding the dispersion found by analysing the {\it Gaia} benchmark stars with iSpec as per  \cite{2014A&A...566A..98B,2014A&A...564A.133J,2015A&A...582A..49H}.

\begin{figure}
	\includegraphics[width=\columnwidth]{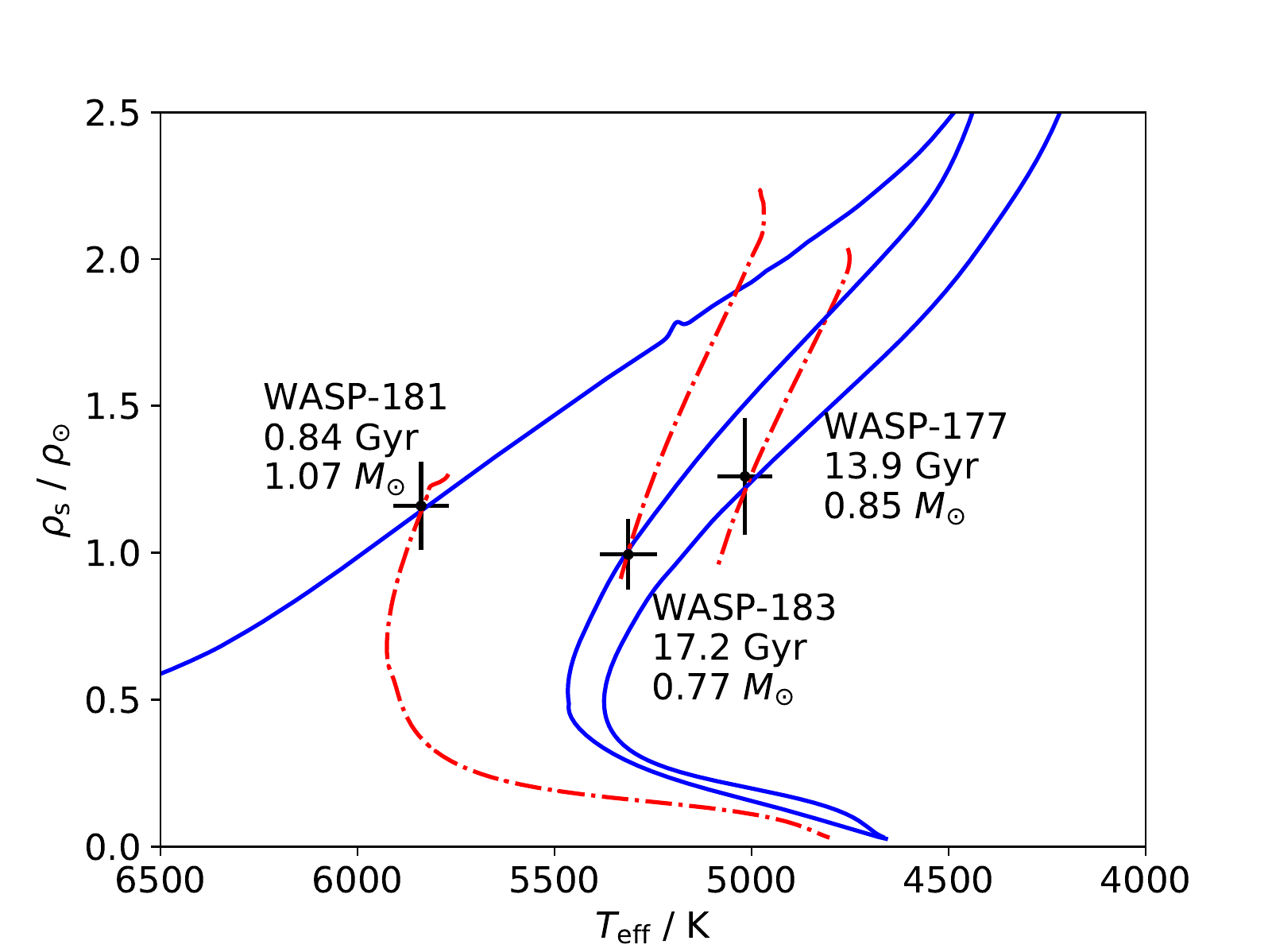}
    \caption{Isochrones (solid/blue) and evolution tracks (dot-dashed/red) output by BAGEMASS for each of the planets we present with the corresponding isochrone age and mass (labelled).}
    \label{fig:tracks}
\end{figure}

We determined the stellar density from and initial fit to the lightcurves and then used it along with the $T_{\rm eff}$ and metallicity to determine stellar masses, for later use in the joint analysis, and the stellar ages with the Bayesian stellar evolution code BAGEMASS \citep{2015A&A...575A..36M}. The resulting parameters are presented in the top part of Table~\ref{tab:all_results} and the corresponding isochrons/evolutionary tracks are shown in Fig~\ref{fig:tracks}. 

BAGEMASS uses an MCMC with a densely sampled grid of stellar models to compute stellar masses and ages. There are three usable grids with differing mixing length parameters, $\alpha_{\rm MLT}$, and helium enhancement.  The default values of these are $\alpha_{\rm MLT} = 1.78$ and no He-enhancement. We used the BAGEMASS default parameters to model WASP-177 and WASP-181 but found that they did not fit WASP-183 very well. This is likely because WASP-183 is among the $\sim 3\%$ of the K-dwarf population that are larger than models would predict \citep{2013ApJ...776...87S}. To account for this we follow the method of \cite{2015A&A...575A..36M} in the case of Qatar-2 and use a the grid provided by BAGEMASS with $\alpha_{\rm MLT} = 1.5$. This results in a much improved fit to the observed density and temperature. We find that the resulting mass estimate is unaffected.

\subsection{System parameters}

To determine the system parameters we modeled the discovery and follow-up data together using the most recent version of the Markov-Chain Monte Carlo (MCMC) code described in detail in \cite{2007MNRAS.380.1230C} and \cite{2015A&A...575A..61A}. We modeled the transit lightcurves using the models of \cite{2002ApJ...580L.171M} with the 4 parameter limb darkening law of \cite{2000A&A...363.1081C,2004A&A...428.1001C}. 

In brief, the models were initialised using the period, $P$, epoch, $T_{0}$, transit depth, $(R_{p}/R_{s})^{2}$, transit duration, $T_{14}$, and impact parameter, $b$, output by the BLS search of each discovery lightcurve. The spectroscopic stellar effective temperature, $T_{\rm eff}$, and metallicity, [Fe/H], were used initially to estimate the stellar mass using the updated Torres mass calibration by \cite{2011MNRAS.417.2166S}. 
To explore the effect of limb-darkening we extracted tables of limb- darkening parameters in each photometric band used for each star. They were extracted for a range of effective temperatures while keeping the stellar metallicity and surface gravity constant. The values used were perturbed during the MCMC via $T_{\rm L-D}$, the `limb-darkening temperature', which has a mean and standard deviation corresponding to the spectroscopic $T_{\rm eff}$ and its uncertainty. 

At each step of the MCMC each of these values are perturbed and the models are re-fit. These new proposed parameters are then accepted if the $\chi^2$ of the fit is better or accepted with a probability proportional to $\exp(-\Delta\chi^2)$ if the $\chi^2$ of the fit is worse.

In the final MCMCs, in place of using the Torres relation to determine a mass, we provided the value given by BAGEMASS. The code then drew values at each step from a Gaussian with a mean and standard deviation given by the value and its uncertainty respectively. Due to the lack of good quality follow-up photometry we imposed a similar prior on the radius of the star WASP-183 using the Gaia parallax. Lacking a complete, good quality follow-up lightcurve can lead to a poor determination of, $\Delta F$, $T_{14}$ and $b$ which we use to calculate the $R_{*}/a$. This in turn results in a poorer determination of $R_{*}$, $R_{p}$ and other parameters that depend upon them. 

In this way we also explored models allowing for eccentric orbits and the potential for linear drifts in the RVs. There was no strong evidence supporting either scenario so we present the system solutions corresponding to circular orbits \citep{2012MNRAS.422.1988A} with no trends due to unseen companions. The parameters derived by these fits can be found in the lower part of Table~\ref{tab:all_results}. 

\subsection{Rotational modulation}

We checked the WASP lightcurves of the three stars for rotational modulation that could be caused by star spots using the method described by \cite{2011PASP..123..547M}. The transits were fit with a simple model and removed. We performed the search over 16384 frequencies ranging from 0
to 1 cycles/day. Due to the limited lifetime and variable distribution of star spots this modulation is not expected to be coherent over long periods of time. As such, we modeled each season of data from each camera individually. WASP-181 and WASP-183 show no significant modulation, with an upper limit on the amplitude of 2- and 3-mmag respectively. 

\begin{figure}
	\includegraphics[width=\columnwidth]{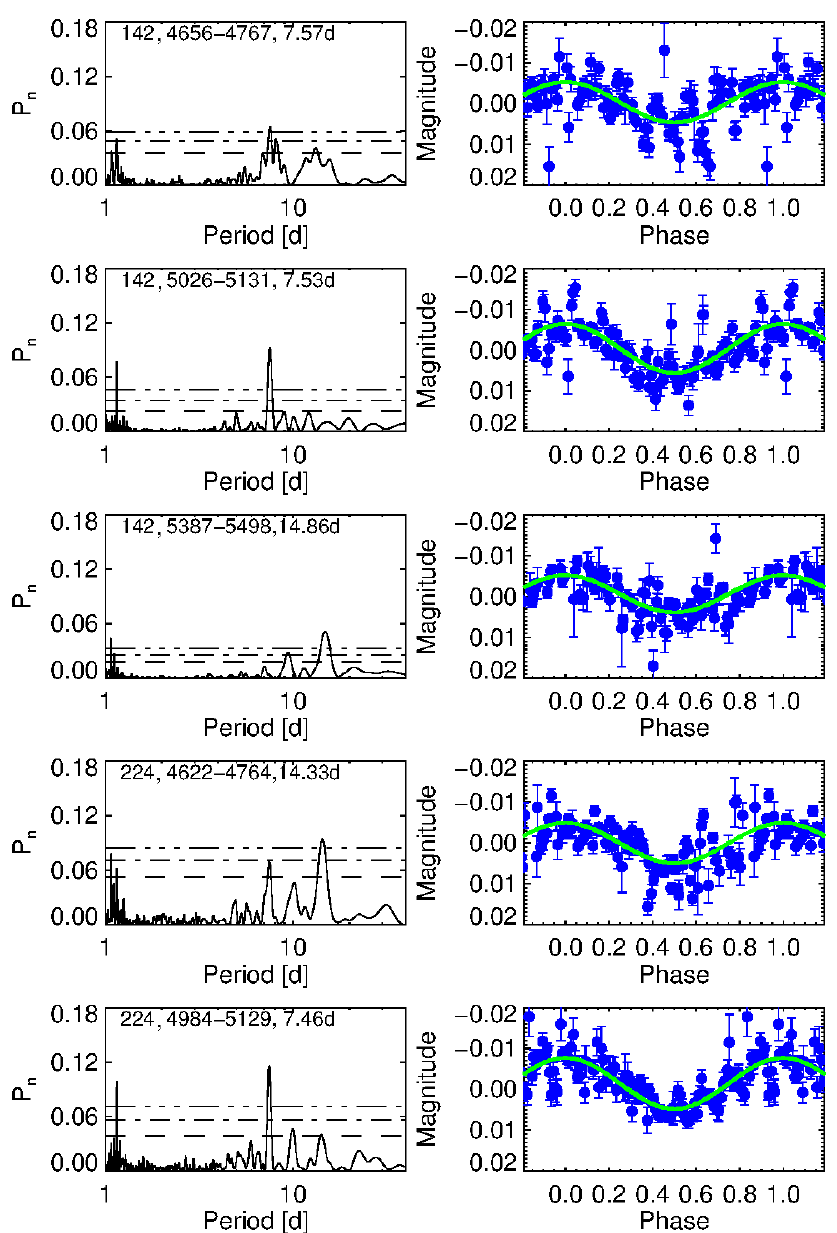}
    \caption{{\it Left}: Periodograms of the WASP lightcurves of WASP-177. Each is labeled with the corresponding camera ID, dates of the observation period (in JD-2450000) and period of the most significant signal. Horizontal lines indicate false-alarm probability levels of 0.1, 0.01 and 0.001. {\it Right}: Lightcurves folded on the most significant detected period.}
    \label{fig:rot-mod}
\end{figure}

However, WASP-177 was found to exhibit modulation consistent with a rotational period, $P_{\rm rot} = 14.86 \pm 0.14$ days and amplitude of $5 \pm 1$ mmag. The results of this analysis for each camera and season of data is shown in Table~\ref{tab:rot_mod}. Fig.~\ref{fig:rot-mod} shows the periodograms of the fits and the discovery lightcurves phase-folded on the corresponding period of modulation. Three of the datasets exhibit $P_{\rm rot} \sim 7$-days while the other two exhibit $P_{\rm rot} \sim 14$-days. We interpret the $\sim 7$-day signals as a harmonic of the longer $\sim 14$-day signal as it is more easy for multiple active regions to produce a $\sim 7$-day signal when the true period is $\sim 14$-days than vice versa. Using this rotational period and our value for the stellar radius we find a stellar rotational velocity of, $v_{*} = 2.9 \pm 0.2$ km/s. When compared to the projected equatorial spin velocity we find a stellar inclination to our line of sight of $38 \pm 25\degr$ which suggests that WASP-177\,b could be quite mis-aligned.

\begin{table}
\caption{Periodogram analysis for WASP lightcurves of WASP-177.}
\label{tab:rot_mod}
\begin{tabular}{llrrrl}\hline
WASP & Dates & Period & Amp & FAP & Notes \\
 Inst. & JD-2450000 & $P_{\rm rot}$ (d) & (mag.) & &  \\
\hline
\\North & 4656-4767 & 7.569 & 0.005 & 0.0017 & P/2\\
North & 5026-5131 & 7.528 & 0.006 & <0.0001 & P/2\\
North & 5387-5498 & 14.860 & 0.004 & <0.0001 & \\
South & 4622-4764 & 14.330 & 0.005 & 0.0007 & \\
South & 4984-5129 & 7.456 & 0.006 & <0.0001 & P/2\\
\hline 
\end{tabular} 
\end{table}

\section{Discussion}

\begin{table*}
\begin{minipage}{\textwidth} 
\caption{System parameters}
\label{tab:all_results}
\begin{tabular}{llccc}\hline
Parameter & Symbol (Unit) & WASP-177 & WASP-181 & WASP-183\\ 
\hline\hline 
\\ 1SWASP ID & $-$ & J221911.19-015004.7 & J014710.37+030759.0 & J105509.36-004413.7 \\ 
Right ascension & (h:m:s) & 22:19:11.19 & 01:47:10.37 & 10:55:09.36 \\ 
Declination & (\degr:':'') & -01:50:04.7 & +03:07:59.0 &  -00:44:13.7 \\
V magnitude & $-$ & 12.58 & 12.91 & 12.76 \\ 
Spectral type\footnote{Spectral type estimated by comparison of $T_{\rm eff}$ to the table in \cite{2008oasp.book.....G}.} & $-$ & K2 & G2 & G9/K0 \\ 
Stellar effective temperature & $T_{\rm eff}$ ($K$) & $5017 \pm 70$ & $5839 \pm 70$ &  $5313 \pm 72$\\ 
Stellar surface gravity & $\log(g)$ (cgs) & $4.49 \pm 0.07$ & $4.38 \pm 0.08$ & $4.25 \pm 0.09$ \\ 
Stellar metallicity & [Fe/H] (dex) &  $0.25 \pm 0.04$ & $0.09 \pm 0.04$ & $-0.31 \pm 0.04$ \\
Projected equatorial spin velocity & $V_{*}\sin I_{*}$ ($km/s$) &  
$1.8 \pm 1.0$ &  $3.3 \pm 0.9$  & $1.0 \pm 1.0$ \\
Stellar macro-turbulent velocity\footnote{Derived via the method of \cite{2014MNRAS.444.3592D}.} & $V_{\rm mac}$ ($km/s$) &  $2.7$ &  $3.3$  & $2.8$ \\
Stellar age & (Gyr)  & $9.7 \pm 3.9$ & $2.5 \pm 1.7$ &  $14.9 \pm 1.7 $  \\
Distance\footnote{From Gaia DR2 \cite{2016A&A...595A...1G,2018A&A...616A...1G,2018A&A...616A...9L}.} &  (pc)  & $178 \pm 2$ & $443 \pm 8$ &  $328 \pm 4$  
\\
\hline
Period & $P$ (d) & $3.071722 \pm 0.000001$ & $4.5195064 \pm 0.0000034$ & $4.1117771 \pm 0.0000051$ \\ 
Transit Epoch & $T_{0} - 2450000$  & $7994.37140 \pm 0.00028$ & $7747.66681 \pm 0.00035$ & $7796.1845 \pm 0.0024$ \\ 
Transit Duration & $T_{14}$ (d) & $0.0672 \pm 0.0013$ & $0.1277 \pm 0.0015$ & $0.084 \pm 0.005$ \\ 
Scaled Semi-major Axis & $a/R_{s}$ &  $9.61^{+0.42}_{-0.53}$ & $12.09 \pm 0.54$ & $11.44 \pm 0.54$ \\ 
Transit Depth & $(R_{p}/R_{s})^{2}$ &  $0.0185^{+0.0035}_{-0.0014}$ & $0.01590 \pm 0.00038$ &  $0.0226^{+0.0060}_{-0.0036}$ \\ 
Impact Parameter & $b$ &  $0.980^{+0.092}_{-0.060}$ &  $0.34^{+0.10}_{-0.15}$ &  $0.916^{+0.163}_{-0.091}$ \\ 
Orbital Inclination & $i$ (\degr) &  $84.14^{+0.66}_{-0.83}$ &  $88.38^{+0.76}_{-0.59}$ &  $85.37^{+0.61}_{-0.88}$ \\ 
Systemic Velocity & $\gamma$ $(\rm kms^{-1})$ & $-7.1434 \pm 0.0041$ & $-8.5489 \pm 0.0072$ & $68.709 \pm 0.012$ \\ 
Semi-amplitude & $K_1 (\rm ms^{-1})$ & $77.3 \pm 5.2$ & $35.7 \pm 3.9$ & $74.8 \pm 6.6$ \\ 
Semi-major Axis & $a$ (au) & $0.03957 \pm 0.00058$ & $0.05427 \pm 0.00069$ & $0.04632 \pm 0.00075$ \\ 
Stellar Mass & $M_{s}$ $(M_{\odot})$ & $0.876 \pm 0.038$ & $1.04 \pm 0.04$ & $0.784 \pm 0.038$ \\ 
Stellar Radius & $R_{s}$ $(R_{\odot})$ & $0.885 \pm 0.046$ & $0.965 \pm 0.043$ & $0.871 \pm 0.038$ \\ 
Stellar Density & $\rho_{s}$ ($\rho_{\odot}$) &  $1.26^{+0.23}_{-0.15}$ & $1.16 \pm 0.15$ & $1.19 \pm 0.17$ \\ 
Stellar Surface Gravity & $\log(g_{s})$ (cgs) &  $4.486^{+0.049}_{-0.037}$ & $4.487 \pm 0.039$ & $4.452 \pm 0.043$ \\ 
Limb-darkening Temperature & $T_{\rm L-D}$ (K) & $5012 \pm 69$ & $5835 \pm 70$ & $5313 \pm 72$ \\ 
Stellar Metallicity & ${[}\textrm{Fe/H}{]}$ & $0 \pm 0$ & $0 \pm 0$ & $0 \pm 0$ \\ 
Planet Mass & $M_{p}$ $(M_{\textrm{Jup}})$ & $0.508 \pm 0.038$ & $0.299 \pm 0.034$ & $0.502 \pm 0.047$ \\ 
Planet Radius & $R_{p}$ $(R_{\textrm{Jup}})$ &  $1.58^{+0.66}_{-0.36}$ &  $1.184^{+0.071}_{-0.059}$ &  $1.47^{+0.94}_{-0.33}$ \\ 
Planet Density & $\rho_{p}$ ($\rho_{\textrm{Jup}})$ &  $0.130^{+0.153}_{-0.085}$ & $0.179 \pm 0.033$ &  $0.16^{+0.18}_{-0.12}$ \\ 
Planet Surface  & $\log(g_{p})$ (cgs) &  $2.67^{+0.22}_{-0.31}$ & $2.686 \pm 0.065$ &  $2.72^{+0.22}_{-0.43}$ \\ 
Planet Equilibrium Temperature\footnote{Assuming 0 albedo and complete redistribution of heat.} & $T_{eq} (K)$ & $1142 \pm 32$ &  $1186^{+32}_{-26}$ & $1111 \pm 30$ \\ 
\\ 
\hline
\end{tabular}
\end{minipage}
\end{table*}

\begin{figure}
	\includegraphics[width=\columnwidth]{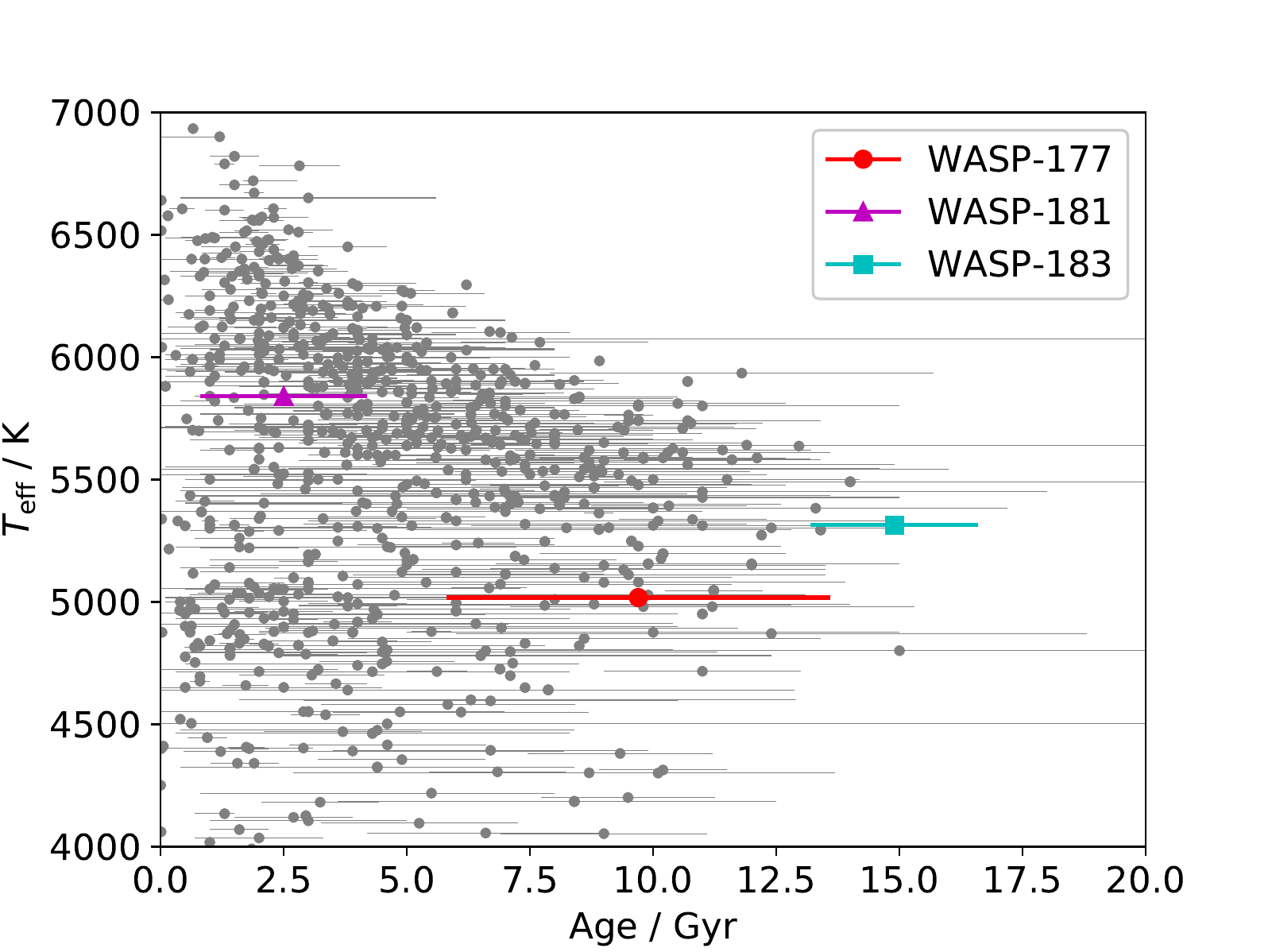}
    \caption{Age distribution for known exoplanet hosts with published uncertainties (grey) and planets presented in this paper (see legend). WASP-183 appears to be particularly old amongst planet hosts. However, we note it is unphysically old and so caution that this determination may be in part due to the K-dwarf radius anomaly. (Data from exoplanet.eu.)}
    \label{fig:age}
\end{figure}

Our joint analysis shows that in this ensemble we have two large sub-Jupiter mass planets: WASP-177\,b ($\sim$0.5 M$_{\rm Jup}$, $\sim$1.6 R$_{\rm Jup}$) and WASP-183\,b ($\sim$0.5 M$_{\rm Jup}$, $\sim$1.5 R$_{\rm Jup}$) orbiting old stars. The third planet, WASP-181\,b, is a large Saturn mass planet ($\sim$0.3 M$_{\rm Jup}$, $\sim$1.2 R$_{\rm Jup}$) . According to the analysis with BAGEMASS, WASP-177 and WASP-183 are both at the latter end of the main sequence explaining their slightly larger radii for stars of their spectral class; a $9.7 \pm 3.9$ Gyr K2 and $14.9 \pm 1.7$ Gyr G9/K0 respectively. WASP-183 is particularly noteworthy as its advanced age makes it one of the oldest stars known to host a transiting planet (see Fig.~\ref{fig:age}). Though, WASP-183 appears to be subject to the K-dwarf radius anomaly, making this determination less clear. Meanwhile, WASP-181 is a relatively young, standard example of a G2 star. 

We compared the stellar radii derived from our MCMC to those we can calculate using the Gaia DR2 parallaxes \citep{2018A&A...616A...9L,2018A&A...616A...1G}, with the correction from \cite{2018ApJ...862...61S}, and stellar angular radii from the infra-red flux fitting method (IRFM) these radii, with reddening accounted for by the use of dust maps \citep{2011ApJ...737..103S}. We find good agreement and present a summary in table \ref{tab:radii_comp}.

\begin{table}
\begin{minipage}{0.45\textwidth}
\caption{Comparison of stellar radii output by the MCMC analysis with radii derived from Gaia DR2.}
\label{tab:radii_comp}
\begin{tabular}{llll}\hline
 Radius source& WASP-177 & WASP-181  & WASP-183\\
\hline
MCMC & $0.885\pm0.046$ & $0.965 \pm 0.043$ & $0.871 \pm 0.038$ \\
Gaia parallax  & $0.80\pm 0.04$ & $0.97 \pm 0.06$ & $0.87 \pm 0.04$ \\
 + IRFM (Corrected)& & & \\
Gaia parallax  & $0.81 \pm 0.04$ & $1.01 \pm 0.05$ & $0.89 \pm 0.04$ \\
 + IRFM (Uncorrected) & & & \\
 \hline
Reddening & 0.072 & 0.023 & 0.04 \\
\hline 
\end{tabular}
\end{minipage}
\end{table}

\begin{figure}
	\includegraphics[width=\columnwidth]{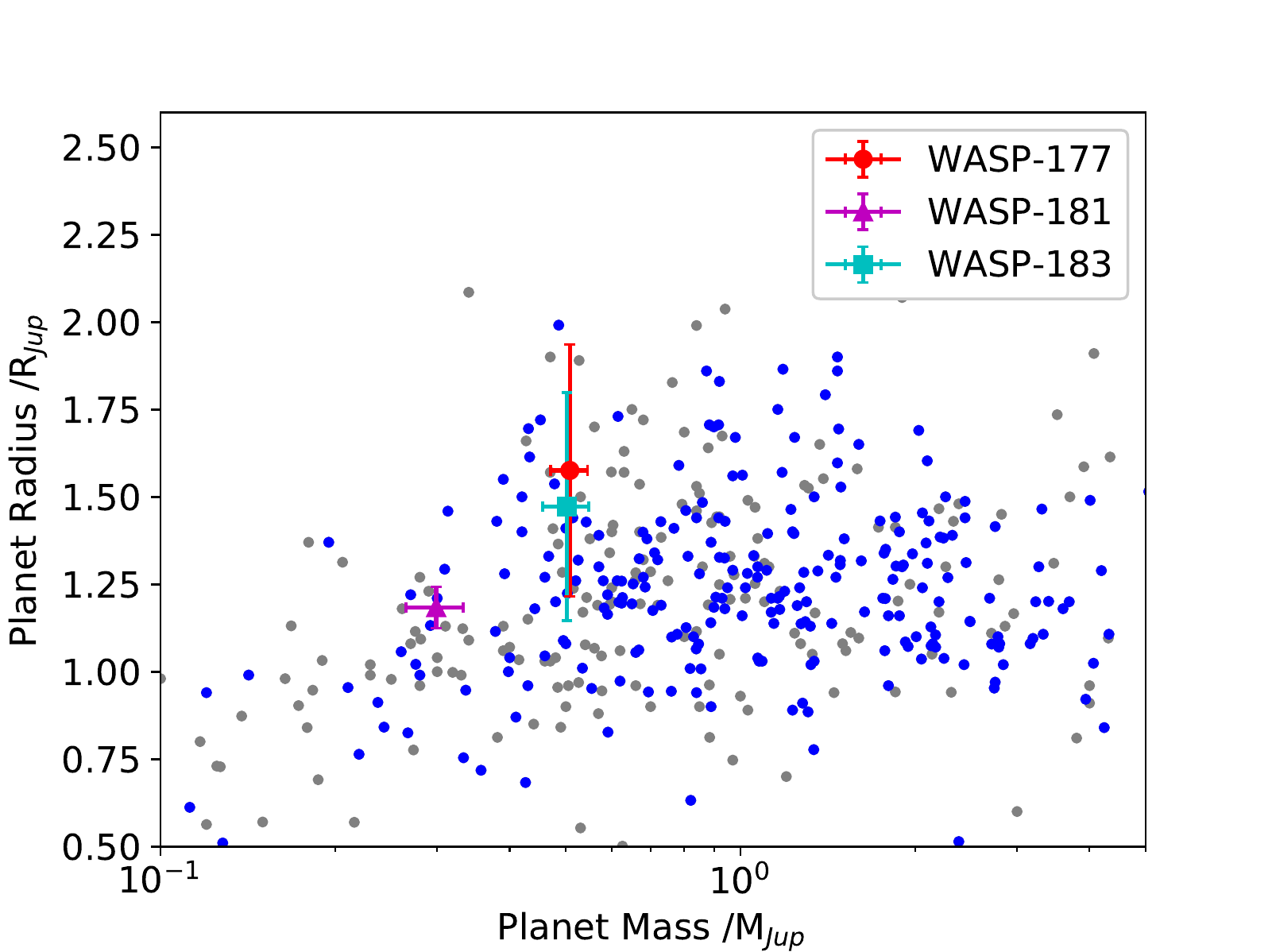}
    \caption{Mass-radius distribution for transiting planets. planets with masses determined to better than 10\% precision are plotted in blue, otherwise the symbols are gray. WASP-177\,b, WASP-181\,b and WASP-183\,b have been plotted with their error bars. Each is close to the upper most part of the distribution. WASP-177\,b is in an area particularly sparsely populated by planets with well determined masses. (Prepared using data collated the TEPCat.)}
    \label{fig:mass-radius}
\end{figure}

All three planets occupy the upper edge of the mass-radius distribution, seen in Fig~\ref{fig:mass-radius}. WASP-181\,b is amongst the group of the largest  planets for an object of its mass. While its mass is not as well determined as the other two, further HARPS observations will help to refine this. WASP-177\,b and WASP-183\,b do lie above the bulk of the distribution, especially when compared to other objects with mass determinations of 10\% precision or better. However, it is difficult to say how exceptional they are as a precise radius determination has proven difficult for them both. The transit of WASP-177\,b is grazing and the transit of WASP-183\,b, in addition to being grazing, lacks a full high precision follow-up lightcurve to refine the transit shape. We anticipate that {\it TESS} observations could soon solve the latter problem;the long cadence data would capture roughly 24 in transit points with a predicted precision from the {\it ticgen} tool of better than 1000 ppm in each 30-minute observation. 

We used the the values derived for planet equilibrium temperature, $T_{\rm eq}$, and surface gravity, $g$, along with Boltzmann's constant, $k$, and the atmospheric mean molecular mass, $\mu$, to estimate the scale heights, $H$, of these planets as:

\begin{equation}
H = \frac{k T_{\rm eq}}{g \mu}
\end{equation}

assuming an isothermal, hydrogen dominated atmosphere. The resulting scale heights were; $790 \pm 320$ km, $770 \pm 200$  km, $696 \pm 464$  km for WASP-177\,b, WASP-181\,b and WASP-183\,b respectively. These translate to transit depth variations of just under 300 ppm for WASP-177 and WASP-181 and $\sim 300$ ppm for WASP-183. If we account for the K-band flux and scale in the same way as \cite{2017A&A...604A.110A}, we get atmospheric signals of; 70, 41 and 60. In reality, we can expect this metric to be an over estimate of detectability for WASP-177\,b and WASP-183\,b as the grazing nature of their transits reduces the impact of the atmospheric signal further. For comparison we used the same metric on other planets with atmospheric detections: water has been detected in the atmospheres of both WASP-12\,b (\citealt{2015ApJ...814...66K}; signal $\sim 93$) and WASP-43\,b (\citealt{2014ApJ...793L..27K}; signal $\sim 74$); titanium oxide has been detected in the atmosphere of WASP-19\,b (\citealt{2017Natur.549..238S}; signal $\sim 83$); sodium and potassium have both been detected in the atmosphere of WASP-103\,b (\citealt{2017A&A...606A..18L}; signal $\sim 37$). While not ideal targets, this suggests such detections may be possible.

Investigation into any eccentricity or long-period massive companions in these systems has not yielded anything convincing. All of the orbits are circular, with the $2\sigma$ upper limits quoted in Table~\ref{tab:all_results}. As for long term trends, WASP-177 shows the possibility of a very low significance ($\sim 1.5 \sigma$) drift with $\delta\gamma/\delta t$ of $(-2.4 \pm 1.6)\times 10^{-5}$ km/s/d. Neither WASP-181 nor WASP-183 show significant drifts with $\delta\gamma/\delta t$ of $(1.2 \pm 4.0)\times 10^{-5}$ km/s/d and $(-1.9 \pm 5.1)\times 10^{-5}$ km/s/d respectively. 

\begin{figure}
	\includegraphics[width=\columnwidth]{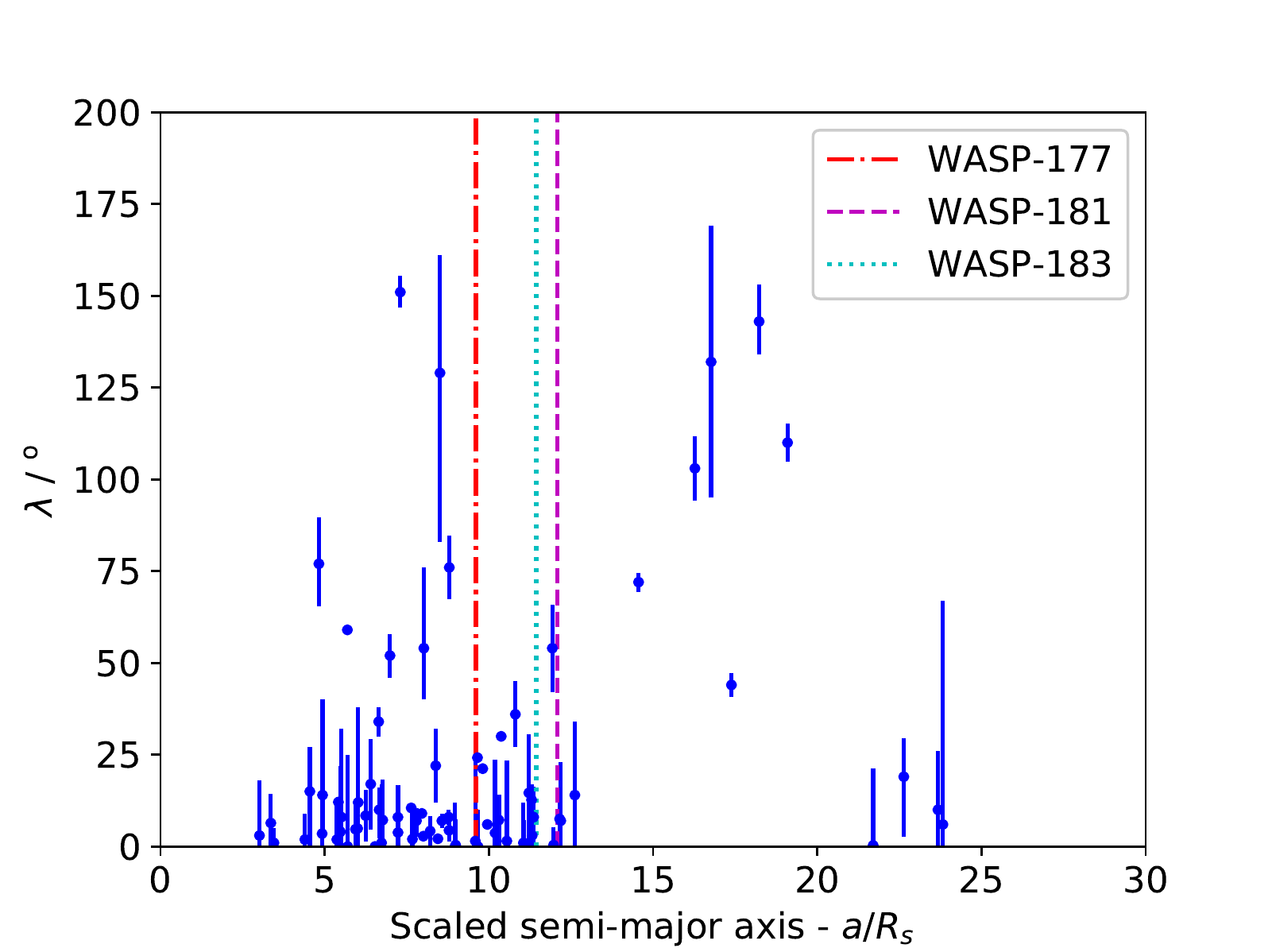}
    \caption{Distribution of planets with measured spin-orbit angles with cool host stars. WASP-177, WASP-181 and WASP-183 are all cool stars by this definition and the planets lie in the region where mis-alignment is often said to become more common. WASP-177 shows signs of being misaligned and so may be an interesting diagnostic in this region.}
    \label{fig:spin-orbit}
\end{figure}

Finally, these systems do present interesting targets for the investigation of the observed spin-orbit mis-alignment distribution (\citealt{2012ApJ...757...18A,2015ApJ...800L...9A}). All of the stellar hosts fall into the "cool" regime of \cite{2012ApJ...757...18A} and despite their short periods have scaled semi-major axes, $a/R_{*}$, above 8. They are therefore above the empirical boundary noted by \cite{2017AJ....153..205D} as the transition region where systems with cooler stars show more tendency to be mis-aligned. Since the study in 2017 the number of systems with obliquity measurements has increased. Most of the cool-star systems with $a/R_{*}$ above 8 are well aligned, see Fig~\ref{fig:spin-orbit}.  

We estimated the alignment time-scale for each system using Eq.4 of \cite{2012ApJ...757...18A} as was done for WASP-117 \citep{2014A&A...568A..81L}. These time-scales, along with the mass of the convective zone, $M_{\rm cz}$, are shown in Tab~\ref{tab:alignment}. In each case, the time-scale for realignment is much longer than the ages of the systems. Therefore, we would expect the initial state of alignment of the systems to have been preserved. We have estimated the inclination of WASP-177 to be $38 \pm 25\degr$ and so may expect it to join only 12 systems with  $a/R_{*}$ < 15 that show mis-alignment this makes it a potentially important diagnostic in determining the factors that cause or preserve mis-alignment. 

\begin{table}
\centering
\begin{minipage}{0.45\textwidth}
\caption{Convective zone masses and estimated time-scales for realignment of systems in this paper.}
\label{tab:alignment}
\begin{tabular}{lrr}\hline
Star & $M_{\rm cz}$\footnote{Dervied from \cite{2001ApJ...556L..59P}.} & $\tau$ \\
 & ($M_{\odot}$) & (Gyr) \\
\hline
\\WASP-177 & $10^{-1.3}$ & 120 \\
WASP-181 & $10^{-1.7}$ & 7500 \\
WASP-183 & $10^{-1.4}$ & 200 \\
\hline 
\end{tabular} 
\end{minipage}
\end{table}

We calculate that the amplitude of the RM effect will be greatest for WASP-181 at $\sim 50$ ms$^{-1}$. The effect should also be detectable for WASP-177 and WASP-183 despite their more grazing transits, with an amplitude of $\sim 10$ ms$^{-1}$.

\section{Conclusions}

We have presented the discovery of 3 transiting exoplanets from the WASP survey; WASP-177\,b ($\sim$0.5 M$_{\rm Jup}$, $\sim$1.6 R$_{\rm Jup}$), WASP-181\,b ($\sim$0.3 M$_{\rm Jup}$, $\sim$1.2 R$_{\rm Jup}$), and WASP-183\,b ($\sim$0.5 M$_{\rm Jup}$, $\sim$1.5 R$_{\rm Jup}$). They all occupy the upper region of the mass-radius distribution for hot gas-giant planets but do not present exceptional targets for transmission spectroscopy. However, regarding the investigation of system spin-orbit alignment they do occupy an under investigated range of $a/R_{*}$ and so could act as good probes of tidal realignment time-scales.

\section*{Acknowledgements}

We thank the Swiss National Science Foundation (SNSF)
and the Geneva University for their continuous support to our planet search programs. This work has been in particular carried out in the frame of the National Centre for Competence in Research `PlanetS' supported by the Swiss National Science Foundation (SNSF). WASP-South is hosted by the South African Astronomical Observatory and we are grateful for their ongoing support and assistance.
Funding for WASP comes from consortium universities and from the UK's Science and Technology Facilities Council. 
TRAPPIST is funded by the Belgian Fund for Scientific Research (Fond National de la Recherche Scientifique, FNRS) under the grant FRFC 2.5.594.09.F, with the participation of the Swiss National Science Fundation (SNF). MG is a F.R.S.-FNRS Senior Research Associate.
The research leading to these results has received funding from the European Research Council under the FP/2007-2013 ERC Grant Agreement 336480, from the ARC grant for Concerted Research Actions  financed  by  the  Wallonia-Brussels  Federation,  from the Balzan Foundation, and a grant from the Erasmus+ International Credit Mobility programme (K Barkaoui). We thank our anonymous reviewer for their comments which helped improve the clarity of the paper.  




\bibliographystyle{mnras}
\bibliography{w177-w181-w183_bib} 


\clearpage

\appendix
We include the data we used in this paper as online material. Examples of the tables are show here.

\section{Online Data}

\begin{table}
\begin{minipage}{\textwidth} 
\caption{Data from WASP}
\label{tab:DATA-WASP}
\begin{tabular}{ccccc}\hline
 BJD -2450000 & Diff.  & Mag. & Target\\
  & magnitude & error & 
\\ 
\hline 
\\5026.54902768 & -0.00254900 & 0.01949100 & WASP-177\\
5026.54946749 & 0.02243000 & 0.01957700 & WASP-177 \\
5026.55550916 & -0.00315500 & 0.01926500 & WASP-177 \\
5026.55596055 & -0.00210900 & 0.01891800 & WASP-177 \\
5026.56091425 & 0.02301800 & 0.01892700 & WASP-177 \\
5026.56135407 & -0.01070600 & 0.01829000 & WASP-177 \\
5026.56629620 & -0.01820900 & 0.01836500 & WASP-177 \\
5026.56673601 & -0.03087100 & 0.01769000 & WASP-177 \\
5026.57268508 & -0.02453400 & 0.01780000 & WASP-177 \\
5026.57312490 & -0.00700800 & 0.01818600 & WASP-177 \\
\\ 
\hline 
\end{tabular} 
\end{minipage}
\end{table} 

\begin{table}
\begin{minipage}{\textwidth} 
\caption{Data from Trappist}
\label{tab:DATA-TRAP}
\begin{tabular}{ccccc}\hline
 BJD -2450000 & Dif. Mag. & Mag. error & Filter & Target\\ 
\hline 
\\7960.51599185 & -0.00760377 & -0.00345472 & I+z & WASP-177 \\
7960.51636185 & -0.00029799 & -0.00344426 & I+z & WASP-177 \\
7960.51664185 & 0.00210904 & -0.00344268 & I+z & WASP-177 \\
7960.51691185 & -0.00560489 & -0.00344076 & I+z & WASP-177 \\
7960.51718185 & -0.00165321 & -0.00342985 & I+z & WASP-177 \\
7960.51754185 & -0.00448940 & -0.00342637 & I+z & WASP-177 \\
7960.51782185 & -0.00682232 & -0.00342797 & I+z & WASP-177 \\
7960.51809185 & 0.00938183 & -0.00343320 & I+z & WASP-177 \\
7960.51836185 & -0.00237813 & -0.00343161 & I+z & WASP-177 \\
7960.51863185 & -0.00132194 & -0.00342244 & I+z & WASP-177 \\
\\ 
\hline 
\end{tabular} 
\end{minipage}
\end{table} 

\begin{table}
\begin{minipage}{\textwidth} 
\caption{RV data}
\label{tab:RV-DATA}
\begin{tabular}{ccccc}\hline
 JD -2450000 & RV & RV error & Instrument & Target\\
 & (km/s) & (km/s) & & 
\\ 
\hline 
\\7626.633110 & -7.19243 & 0.01963 & CORALIE & WASP-177 \\
7629.687997 &  -7.21044 &  0.03748 & CORALIE & WASP-177 \\
7689.581199 &  -7.05873 &  0.01634 & CORALIE & WASP-177 \\
7695.567558 &  -7.10068 &  0.01812 & CORALIE & WASP-177 \\
7933.845373 &  -7.16482 &  0.02686 & CORALIE & WASP-177 \\
7937.771917 &  -7.12978 &  0.02180 & CORALIE & WASP-177 \\
7952.880188 &  -7.14280 &  0.02144 & CORALIE & WASP-177 \\
7954.787481 &  -7.19749 &  0.01473 & CORALIE & WASP-177 \\
7961.703754 &  -7.19347 &  0.02763 & CORALIE & WASP-177 \\
8047.604223 &  -7.20369 &  0.01660 & CORALIE & WASP-177 \\
\\ 
\hline 
\end{tabular} 
\end{minipage}
\end{table} 

\begin{table*}
\begin{minipage}{\textwidth} 
\caption{Data from Euler}
\label{tab:DATA-EULER}
\begin{tabular}{ccccccccccc}\hline
BJD - 2450000 & Dif. Mag. & Mag. & X-pos & Y-pos & Airmass & FWHM & Sky Bkg. & Exp. time  &  Filter & Object\\
 &   & error & (pix) & (pix) &  & (pix) &  & (s) & (days) &  
\\ 
\hline 
\\8046.53876846 & 0.00034578 & 0.00359381 & 1070.950 & 571.822 & 1.1298 & 9.369 & 0.869 & 110 & B & WASP-177 \\
8046.54029585 & 0.00037222 & 0.00358208 & 1086.396 & 562.842 & 1.1289 & 7.076 & 0.903 & 110 & B & WASP-177 \\
8046.54281198 & -0.8042 & 0.00213508 & 1085.203 & 562.140 & 1.1279 & 7.496 & 2.5836 & 300 & B & WASP-177 \\
8046.54652188 & -0.00024034 & 0.00213411 & 1086.544 & 558.005 & 1.1267 & 7.632 & 2.4149 & 300 & B & WASP-177 \\
8046.55014519 & 0.00031779 & 0.00213539 & 1086.429 & 558.463 & 1.1262 & 7.980 & 2.3836 & 300 & B & WASP-177 \\
8046.55443872 & 0.00313122 & 0.00184258 & 1085.948 & 555.787 & 1.1264 & 7.832 & 3.2365 & 400 & B & WASP-177 \\
8046.55920903 & 0.00363211 & 0.00184463 & 1084.985 & 556.119 & 1.1274 & 7.832 & 3.4133 & 400 & B & WASP-177 \\
8046.56407976 & 0.00279818 & 0.00185139 & 1087.955 & 557.254 & 1.1296 & 7.928 & 3.4343 & 400 & B & WASP-177 \\
8046.56884813 & 0.00639643 & 0.00186126 & 1087.783 & 558.089 & 1.1326 & 9.099 & 3.9564 & 400 & B & WASP-177 \\
8046.57371742 & 0.00938502 & 0.00185610 & 1089.022 & 557.002 & 1.1369 & 7.880 & 3.5795 & 400 & B & WASP-177 \\
\\ 
\hline 
\end{tabular} 
\end{minipage}
\end{table*}



\bsp	
\label{lastpage}
\end{document}